\documentclass[english]{article}
\usepackage[T1]{fontenc}
\usepackage[latin9]{inputenc}
\usepackage{babel}
\begin{document}

\title{Partial Dynamical Symmetries in the f$_{7/2}$ and g$_{9/2}$Shells}

\author{Larry Zamick }

\date{Department of Physics and Astronomy, Rutgers University, Piscataway,
New Jersey 08854}
\maketitle
\begin{abstract}
Previous work on partial dynamical symmetries in the f$_{7/2}$ shell
is extended to other shells e.g. g$_{9/2}$ .The nuclei involved are
$^{43}$Sc ,$^{43}$Ti,$^{44}$Ti, $^{52}$Fe,$^{53}$ Fe, $^{53}$Co,$^{96}$Cd,$^{97}$Cd
and$^{97}$In
\end{abstract}
Previously we found partial dynamical symmetries (PDS) in the f$_{7/2}$
shell when ,in a single j shell model calculation we set all T=0 two
body matrix elements to zero{[}1,2,3,4,5{]}. We found that for selected
angular momenta Jp and Jn were separately good quantum numbers-not
just total J.. Further states with a given ( Jp.Jn) were all degenerate
in energy for these selected states.

It took a while but it was finally realized that the selected states
were those with angular momenta that could not occur for a system
of identical particles. For example in $^{43}$Ca there are no states
in the f$_{7/2}$ model space with angular momenta J=1/2 and J=13/2
Therefore states in $^{43}$Sc with these total J values are degenerate
and have as good quantum numbers (Jp=4 , Jn= 7/2). A similar story
for J=17/2 and 19/2. Here the good quantum numbers are (Jp=6 Jn=7/2).
In $^{44}$Ca there are no states in the f$_{7/2}$ configuration
with J=3,7,9,10,11,12.. States in $^{44}$Ti and $^{52}$ Fe with
Jp=4 Jn=6 and Jp=6 Jn=4 with angular momenta J=3$_{2}$,7$_{2}$,9$_{1}$
and 10$_{1}$ are all degenerate. Also Jn=6 Jp=6 J=10$_{2}$ and 12$_{1}$.The
12+ state is isomeric in $^{44}$Ti bt even more so $^{52}$Fe . In
the later nucleus the 12+ state is is at a lower energy than the 10+
state. In $^{44}$Ti it is a bit above.

We can carry over the arguments to the 9/2 shell. Starting from a
core with Z=50 N=50 the obvious analogs to nuclei in the f$_{7/2}$
shell are $^{97}$Cd ($^{97}$In) and $^{96}$Cd, three holes and
four holes respectively. We stay away from particles added to an N=40
Z=40 core bec ause $^{80}$Zr is not a good closed shell.

In our previous papers we actually gave formulas for general j, not
just j=7/2.

There are two conditions which lead to a PDS. The off diagonal condition
insures that J$_{p}$ and J$_{n}$ are separately good quantum numbers-not
just total J. Then there is the diagonal condition that expains why
states with the same (J$_{p}$,J$_{n}$) are degenerate.

Three particles-off diagonal condition:

$\left\{ \begin{array}{ccc}
j & j & (2j-3)\\
(3j-4) & j & (2j-1)
\end{array}\right\} =0$

Three particles diagonal condition:

$\left\{ \begin{array}{ccc}
j & j & (2j-1)\\
j & J & (2j-1)
\end{array}\right\} $ = (-1)$^{(J+j)}$/(8j-2) for J= (3j-1), (3j-2) and (3j-4).

Four particles off-diagonal condition :

$\left\{ \begin{array}{ccc}
j & j & (2j-1)\\
j & j & (2j-1)\\
(2j-1) & (2j-3) & (4j-4)
\end{array}\right\} =0$

Four particles diagonal condition:

$\left\{ \begin{array}{ccc}
j & j & (2j-3)\\
j & j & (2j-1)\\
(2j-3) & (2j-1) & J
\end{array}\right\} $= 1/{[}4(4j-5) (4j-1){]} for selected J values.

This topic is also of interest in terms of what we call companion
problems{[}6{]}. Initially Shadow Robinson and I used

Regge 6j symmetries to show why certain 6j symbols vanished {[}7{]}.
But ther are no Regge relations for 9j symbols. But then we found
that Talmi {[}8{]} had shown for a completely different reason why
the same 6j vanished. He constructed a coefficient of fractional parentage
for a state with an angular momentum which did not exist for a system
of three identical particles(J=13/2 in $^{43}$Ca).The vanishing of
the cfp was carried by the same 6j symbol we needed to explain the
vanshing of off diagonal coupling for our PDS. We then used these
ideas for 9j symbols. For a 4 nucleon system wecalculate cfp's for
states that do not exist.

In another direction with regards to companion problems Zhao and Arima
{[}9{]}obtained the same 9j relations that we had by considering J
pairing Hamiltonians. It is quite fascinating that quite different
physical problems lead to the same mathematical relations.

Let us look at proceed systematically.For three identical particles
in a j shell the maximum J is j + (j-1)+(j-2) = (3j-3). For one proton
and 2 neutrons the maximum value is (2j-1)+j =(3j-1).. Hence states
with J=3j-2 and 3j-1 are part of the PDS These have high spins and
so the single j model might work better. Also belonging to the PDS.
are states with J=1/2 and 3j-4 The last one belongs because there
are no states with J=J(max) -1 for identical fermions (also true for
identical bosons).

For 4 nucleons (or holes) the maximum J is j+(j-1)+(j-2)+(j-3)=4j-6.
However for two protons and 2 neutrons the maximum J is (2j-1) +(2j-1)=4j-2.
Hence states with J= (4j-5), (4j-4), (4j-3) and (4j-2) belong to the
PDS.. These are high spin states. The single j shell might work fairly
well for these. There might be other states with PDS. e.g. as noted
above . J=3 and 7 in the f$_{7/2}$ shell and J=11 in the g$_{9/2}$shell.

Consider first 3 nucleons in the g$_{9/2}$ shell. If they are identical
J$_{max}$= 21/2. For one proton and 2 neuterons and/or

2 protons and one neutron J$_{max}$= 25/2. We get a degenerate set
J$_{p}$=8 J$_{n}$=9/2 $\qquad$J=19/2, 23/2 and 25/2 (all T=1/2).

Consider four nucleouns in the g$_{9/2}$ shell. If they are identical
J$_{max}$ =12. For 2protons and 2 neuterons J$_{max}$= 16.

Here are selected sets of degenerate states for four nucleons in the
g$_{9/2}$ shell.

$\qquad$J$_{p}$$\qquad$J$_{n}$

$\qquad8$ $\qquad$8 $\qquad$J=14,16$\qquad$ T=0 

$\qquad8$ $\qquad$6 $\qquad$ J=11, 13, 14  T=0 

There are more . In the above (8,6) is an abreviation for (8,6) +
(-1)$^{(J+T)}$(6,8).

(For the (8,6) configuration there is also a degeneracy of J=8 and
J=9. The above considerations do not explain this .)

The T=0 =0 calculation is a good starting point to see the effects
of putting back the T=0 matrix elements. In the f$_{7/2}$ shell.
There was some striking behavior for T=1/2 states of a three particle
system.In a complete fp calculation we considered the difference E(full)
\textendash{} E(T=0=0). The behaviour for J <.7/2 was different from
that for J>7/2. In the former case for deceasing J the above quantity
became increasingly and linearly negatives. For the higher spins these
was as staggering effect with the alternate spins J=9/2, 13/2 ad 17/2
going up in energy when both T=0 and T=1 matrix elements were included
while J=7/2 ,J=11/2 , 15/2 and 19/2 hardly changed. We should expect
similar behaviour in higher j shells. For T=0 states of a four nucleon
system we found that the odd spin states were pushed up significantly
more than even spin when the full interaction was reintroduced. Thus
with only the E(T=0=0) interaction there were several odd spin states
close to or below the lowest J=12+ state. These were pushed up by
the full interaction.

There have been several shell model calculations in the g$_{9/2}$
region including early calculations by Auerbach and

Talmi {[}10{]} , Serduke,Lawson and Gloeckner {[}11{]}and Ogawa {[}12{]}.
The phrase {}``spin gap isomers is used'' and Ogawa

predicted many such isomers in this region $^{95}$Pd, $^{95}$Ag,$^{96}$Cd,$^{97}$Cd.
Reintroducing the T=0 two- body matrix elements

clerly helps to create these spin gaps.

At this workshop (Nuclear Physics in AstrophysicsV) new results on
$^{96}$Cd have been presented. In particular his York

group found a J=16+ isomeric state. For this state to be isomeric
the J=14+ state should be at a higher energy than the

J=16+ state. As note above without the T=0 interaction in the single
j shell calculation these two states would be

degenerate in energy.The T=0 interaction is required to remove the
degneracy and push the J=14+ state above the J=16+

state.

Th re are other PDS in a single j shell calculation for $^{96}$Cd.In
general seniority is not a good quantum number in the

g$_{9/2}$ shell. However we previousy found {[}10{]} that for a system
of four identical particles (holes) with configuration(g$_{9/2})^{4}$
there is a certain J=4 v=4 (aso J=6) state that does not mix with
the other two states,the latter having seniorities v=2 and 4. In other
words there is one eigenfunction for J=4 v=4 which emerges no matter
what interaction is used. This problem has been studied in several
references {[}13{]},{[}14{]},{[}15{]}, and {[}16{]}. Different, and
very wide ranging topics involving partial dynamical symmetries have
been develped and reviewed by A. Leviatan {[}17{]}.

As seen in the references these works were done with several collaborators
Shadow Robinson (Phd thesis), Alberto Escuderos , Ben Bayman and Piet
Van Isacker{[}1,2,3,4,5,15{]}. They were strongly influenced by works
of Igal Talmi{[}8{]}. 

This work was stimulated in part by reports at the Weizmann Institute
workshop following the Eilat conference of relevant experiments in
the g$_{9/2}$ shell by the York group {[}19{]}. As reported by B.S.
Nara Singh A J=16+ isomer was found 1n $^{96}$Cd which decayed to
a J=15+ isomer in $^{96}$Ag {[}19{]}. In the light of his comments
I re-examined work I had done with Escuderos {[}13{]} where the main
thrust was not isomerism but symmetries. But as a residual we did
obtain an isomerism of J=15+ in $^{96}$Ag in a single j-shell calculation
- g$_{9/2}$- with a Q.Q interaction. The energies of the J=15+,14+,13+,12+,11+
states relative to a J=1+ ground state are respectively 2.48, 3.09,
2.53, 2.59 and 1.96 MeV. Thus the J=12+,13+,and 14+ are higher in
energy than J=15+. The latter state can only decay to J=11+ via an
E(4) and/or M(5) transition. Amusingly we find that the J= 8+ state
is also isomeric in this model space. 

. Thanks go to the Weizmann Institute for my support via a Morris
Belkin visiting professor appointment.Help from Justin Farischon is
greatfully acknowleged.

{[}1{]} S.J.Q. Robinson and L. Zamick, Phys. Rev. C 63, 064316 (2001)

{[}2{]} S.J.Q. Robinson and L. Zamick , Phys. Rev. C 64, 057302 (2001)

{[}3{]} S.J.Q. Robinson, A. Escuderos and L.Zamick , Phys.Rev. C 72,
034314 (2005)

{[}4{]} A. Escuderos, B.F. Bayman and L. Zamick , Phys.Rev. C 72,
054361 (2005)

{[}5{]} A. Escuderos, S.J.Q Robinson and L. Zamick, Phys.Rev. C 73,
027301 (2006)

{[}6{]} L. Zamick@ A. Escuderos, Phys. Rev. C 71, 014315 (2005)

{[}7 {]} T. Regge, Il Nuovo Cimento. Vol XI, N.1 (1959) 298

{[}8{]}. Talmi, Simple Models of Complex Nuclei ,Harwood Academic
Publishing Switzerland (1993)

{[}9{]} Y.M. Zhao and A. Arima , Phys. Rev. C 72, 054307 (2005)

{[}10{]} N. Auerbach and I. Talmi, Phys. Let. 10, 297 (1964)

{[}11{]} F.J.D. Serduke,R.D. Lawson and D.H. Gloeckner, Nucl. Phys.
A256,45 (1976)

{]}12{]} K. Ogawa, Phys. Rev C 28, 958 (1983)

{[}13{]} A. Escuderos and L. Zamick Phys. Rev. C73, 044302 (2006)

{[}14{]} L. Zamick Phys. Rev. C75, 064305 (2007)

{[}15{]} P. Van Isacker and S. Heinze, Phys. Rev. Lett 100, 052501
(2008)

{[}16{]} L. Zamick and P. Van Isacker, Phys. Rev. C 78, 044327 (2008).

{[}17{]} Chong Qi, Phys. Rev. C 83,014307 (2011)

{[}18{]} A. Leviatan, Prog. Part. Nucl. Phys. 66,93-143 (2011) 

{[}19{]} B.S. Nara Singh, N\textasciitilde{}Z Physics Around A\textasciitilde{}100,Nuclear
Physics in Astrophysics V-post conference workshop,

$\qquad$Weizmann Institute,April 10-11 (2001)
\end{document}